\documentclass[aps,prl,twocolumn,groupedaddress]{revtex4-1}

\usepackage{graphicx}
\usepackage{hyperref}
\hypersetup{
    colorlinks,
    linkcolor = [RGB]{34,30,128},
    citecolor = [RGB]{34,30,128},
    urlcolor = [RGB]{34,30,128}
}
\begin{document}

\title{Role of Electric Fields on Enhanced Electron Correlation 
in Surface-Doped FeSe}

\author{Young Woo Choi and Hyoung Joon Choi}
\email{h.j.choi@yonsei.ac.kr}
\affiliation{Department of Physics, Yonsei University, Seoul 03722, Republic of Korea}

\date{\today}

\begin{abstract}
Electron-doped high-$T_c$ FeSe reportedly has a strong electron correlation 
that is enhanced with doping. 
It has been 
noticed that significant electric fields exist inevitably between FeSe and 
external donors along with electron transfer. However, the effects of such 
fields on electron correlation are yet to be explored. Here we study 
potassium- (K-) dosed FeSe layers using density-functional theory 
combined with dynamical mean-field theory to investigate the roles of such 
electric fields on the strength of the electron correlation. We find, very 
interestingly, the electronic potential-energy difference between the topmost 
Se and Fe atomic layers, generated by local electric fields of ionized K 
atoms, weakens the Se-mediated hopping between Fe $d$ orbitals. Since it is 
the dominant hopping channel in FeSe, its reduction narrows the Fe $d$ 
bands near the Fermi level, enhancing the electron correlation. This effect 
is orbital dependent and occurs in the topmost FeSe layer only. We also 
find the K dosing may increase the Se height, enhancing the electron 
correlation further. These results shed new light on the comprehensive 
study of high-$T_c$ FeSe and other low-dimensional systems.
\end{abstract}

% insert suggested PACS numbers in braces on next line
% \pacs{}
% insert suggested keywords - APS authors don't need to do this
% \keywords{}

\maketitle

Observations of a superconducting $T_c$ as high as 100 K in a monolayer (ML)
FeSe/SrTiO$_3$ (STO) system \cite{QingYan:2012,He:2013,Ge:2015}
have intensified interest in electron-doped FeSe 
systems \cite{Liu:2015,Huang:2017,Wang:2017}. In the FeSe/STO system, electron 
transfer from the STO substrate to FeSe appears to be a key ingredient for 
realizing superconductivity \cite{He:2013,Tan:2013}, and additional 
electron doping to the system by potassium (K) dosing increases 
$T_c$ \cite{Shi:2017}. Moreover, surface doping experiments 
by K dosing \cite{Tang:2015,Miyata:2015,Wen:2016,Song:2016}, Na dosing
\cite{Seo:2016}, and liquid gating \cite{Lei:2016}
have shown that electron doping can also increase $T_c$ 
for bulk, thick-film, and multilayer FeSe.

Along with the enhanced $T_c$, a common feature shared by electron-doped 
FeSe systems is that they have much stronger electron correlation than iron 
pnictides \cite{Wen:2016,Seo:2016,Yang:2009,Qazilbash:2009,Yi:2015,He:2014}. 
An insulator-superconductor transition was reported in FeSe/STO, which is 
electron doped interfacially, suggesting strong electron 
correlation \cite{He:2014}. Strong renormalization of $d_{xy}$ bands was 
observed, suggesting orbital-dependent electron correlation \cite{Yi:2015}.
Systematic doping experiments with K and Na dosing showed that 
the correlation strength increases with the doping 
level \cite{Seo:2016,Wen:2016}, which is quite anomalous because the 
electron correlation usually decreases with deviation from 3$d^5$ in 
Fe-based superconductors \cite{Wen:2016,Medici:2014,Georges:2013,Nakajima:2014}.

An interesting feature is that the electron doping to FeSe is 
induced usually by 
charge transfer from external donors such as substrates
\cite{Mandal:2017,Tan:2013,Bang:2013,Zou:2016,Zhang:2017} or dosed 
alkali metals \cite{Tang:2015,Miyata:2015,Wen:2016,Seo:2016}. 
In the viewpoint of electrostatics,
such charge transfer is always accompanied by significant 
electric fields between FeSe and external donors.
Thus, not only the doped 
electrons themselves but also some perturbations from the external donors 
may possibly 
affect the electronic structure in FeSe. 
The presence of
such electric fields was noticed
previously in the density-functional theory (DFT) calculation \cite{Zheng:2013,Zheng:2016}; however,
their possible importance on the electron correlation is
not addressed in any previous theoretical or experimental study.

To study the effects of electric fields from external donors,
the simplest prototypical system is K-dosed FeSe.
In our present work, we consider K-dosed ML and bilayer (BL) 
FeSe using DFT combined with the dynamical mean-field theory 
(DMFT) \cite{Kotliar:2006}. 
We obtain that K dosing induces electron doping 
only in the first FeSe layer, and K ions generate a strong local 
electric field near the surface as in other systems \cite{Kim:2015,Baik:2015}.
We show, for the first time, that the electric field 
weakens 
Se-mediated hopping between Fe $3d$ orbitals, reducing DFT bandwidths of
Fe $3d$ 
near the Fermi level and thereby enhancing 
the electron correlation. Effects of the
electric field are mostly contained in the first FeSe layer, 
so the second layer is nearly unaffected.
These effects, caused by electric fields
present generally with external donors,
can be ubiquitous in externally electron-doped FeSe, and may happen
in other low-dimensional materials.
Furthermore, we find the K dosing can increase the Se height 
($z_{\text{Se}}$) from the Fe plane, making FeSe layers more correlated.

\begin{figure} 
\includegraphics[width=8.6cm]{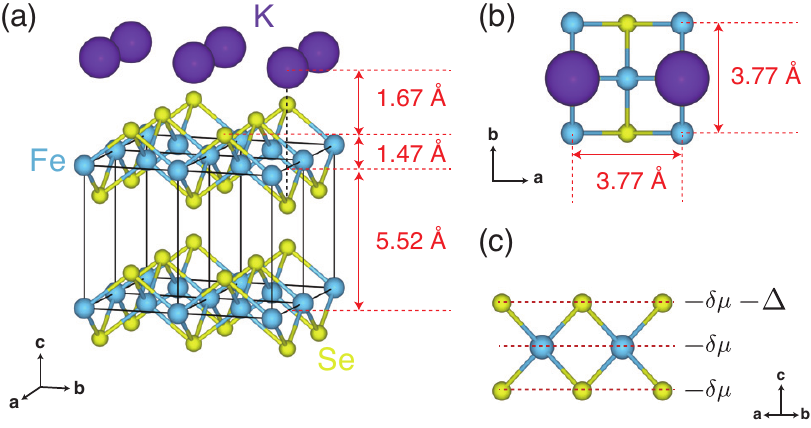}
\caption{\label{fig1} 
K-dosed FeSe layer in (a) perspective and (b) top view, where 
$a=3.77$ {\AA} is from experiment of tetragonal bulk FeSe.
The optimized Se height is 1.47 {\AA} for pristine FeSe.
K atoms are relaxed while Se atoms are fixed as 
the pristine case.  
(c) Schematics of effects of K dosing, where
$\delta\mu$ is the overall potential-energy shift in the first FeSe layer and
$\Delta$ is additional potential-energy change at topmost Se sites.
}
\end{figure}

We performed DFT+DMFT calculations using the all-electron embedded 
DMFT implementation \cite{Haule:2010}, based on WIEN2k \cite{Blaha:2001}. 
The electron correlation in Fe $3d$ orbitals is treated within the DMFT,
whose validity was well tested for 
iron pnictides and chalcogenides \cite{Semon:2017}.
The total electron density is determined using the DFT+DMFT charge self-consistency.
In the DFT part, we use the local density approximation (LDA) to the 
exchange-correlation energy \cite{Perdew:1992}, and $16 \times 16 \times 1$ 
k points are sampled in the full Brillouin zone of the unit cell 
containing two Fe atoms in each FeSe layer, which we call the 2-Fe unit cell 
hereafter. In the DMFT part, we employ the continuous-time quantum Monte 
Carlo impurity solver \cite{Haule:2007} to obtain the local 
self-energies for the Fe $3d$ orbitals, using $U$ = 5.0~eV and $J$ = 
0.8~eV \cite{Yin:2011}. These values of $U$ and $J$, obtained by the 
self-consistent $GW$ method \cite{Kutepov:2010}, have been successful in 
describing various properties of iron pnictides and chalcogenides
\cite{Yin:2011}.
We use the nominal double counting correction scheme and the temperature 
of 116~K.

The atomic structure of FeSe layers is shown in 
Fig.~\hyperref[fig1]{1}. Lattice constants are fixed to experimental 
values of bulk FeSe in the tetragonal phase [Fig.~\hyperref[fig1]{1(b)}]. 
Then, the chalcogen height is the key structural parameter \cite{Moon:2010}.
When relaxed with LDA, $z_{\text{Se}}$ converges to 1.28~{\AA} for bulk, 
much smaller than 1.47~{\AA} in experiment \cite{McQueen:2009}. 
This discrepancy can be resolved using the
DFT+DMFT structural optimization \cite{Haule:2016}. 
By minimizing the DFT+DMFT free energy \cite{Haule:2016}, 
we obtain $z_{\text{Se}} = 1.46$~{\AA} for bulk paramagnetic (PM) phase, 
in good agreement with the experiment \cite{McQueen:2009}
and previous 
calculations \cite{Yin:2011,Haule:2016}. This improvement is related to
fluctuation of local magnetic moments in the PM phase. For FeSe ML, the 
optimized $z_{\text{Se}}$ slightly increases to 1.47~{\AA}. For FeSe BL, 
the distance between Fe layers is set to the experimental value, 
5.52~{\AA}, of bulk FeSe.

\begin{figure} 
\includegraphics[width=8.6cm]{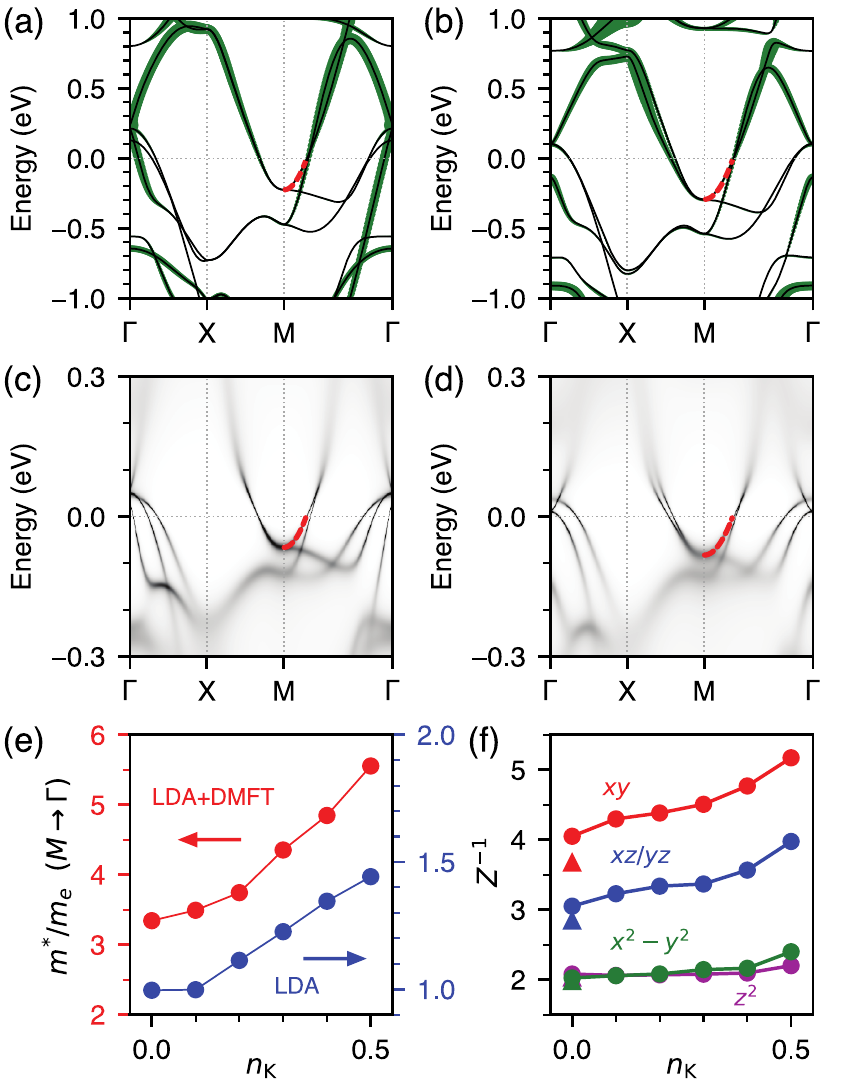}
\caption{\label{fig2} Electronic structure of FeSe ML.
(a),(b) LDA bands (a) without K dosing and (b) with $n_{\text{K}}=0.3$.
Red dashed lines are quadratic fits to bands along $M\rightarrow$$\Gamma$.
The thickness of green solid lines represents Se atomic characters 
of each state. (c),(d) LDA+DMFT spectral functions (c) without K dosing
and (d) with $n_{\text{K}}=0.3$. Red dashed lines are quadratic fits to 
maximal points of spectral functions. (e) Effective mass of the electron 
band at $M$. (f) Inverse of the quasiparticle weights $Z$ for Fe $3d$ orbitals. 
$Z^{-1}$ represents the mass enhancement due to electron correlation.
Triangles are $Z^{-1}$ in bulk FeSe.}
\end{figure}

After determining the atomic positions of pristine FeSe as described 
above, we introduce one K atom per 2-Fe unit cell 
[Figs.~\hyperref[fig1]{1(a)} and \hyperref[fig1]{1(b)}], and optimize the 
K-atom height ($z_{\text{K}}$). 
We define this surface concentration ($n_{\text{K}}$) of K as unity, 
that is, $n_{\text{K}} = 1$. Then, we reduce $n_{\text{K}}$ using the virtual 
crystal approximation (VCA), where the atomic number of K is reduced 
to $18+n_{\text{K}}$ for $n_{\text{K}} > 0$,
which is suitable for describing K atoms randomly 
distributed on the surface as observed experimentally \cite{Song:2016}.
For the pristine case of $n_{\text{K}}=0$, we simply do not introduce 
any K atom. 
We checked the validity of our VCA carefully by comparing it with 
ordinary supercell calculations (See the Supplemental Material \cite{supplemental} 
for the detailed comparison). 
In most of the following parts of our present work, we will focus 
on direct electrical effects of the K dosing on the electron correlation by 
fixing $z_{\text{Se}}$ = 1.47~{\AA} independently of $n_{\text{K}}$. 
Then, at the last part, we will consider the dependence of $z_{\text{Se}}$ on
$n_{\text{K}}$ \cite{note} and its effect on the electron correlation.

First, we investigate charge transfer and the electrostatic potential
after K dosing on the FeSe bilayer. Our LDA results (Fig.~S2 \cite{supplemental}) 
show that the transferred charge from K is contained within the first 
FeSe layer, and the strong electric field appears due to ionized 
K atoms, which is screened by the first FeSe layer, leaving the remaining second 
FeSe layer almost unaffected because each FeSe layer is metallic. 
It can be generalized to any multilayer (see Fig.~S3 \cite{supplemental} 
for four layer). 
Thus, the essential effects of K dosing 
are (i) potential-energy lowering by $\delta\mu$ at the topmost 
FeSe layer with respect to the chemical potential to accommodate electrons 
from K and (ii) additional potential-energy lowering by $\Delta$ at 
the topmost Se atoms with respect to underlying Fe atoms 
[Fig.~\hyperref[fig1]{1(c)}]. We will use these two parameters, $\delta\mu$ 
and $\Delta$, to analyze the electronic structure of K-dosed FeSe layers.

Figures~\hyperref[fig2]{2(a)} and \hyperref[fig2]{2(b)} show our LDA band 
structures of pristine and K-dosed (with $n_{\text{K}} = 0.3$) FeSe MLs, 
respectively. We notice that in the K-dosed ML, bands are shifted downward 
in energy due to electron doping from the K atom, and Fe $3d$ bands near 
the Fermi level are narrower than those in the pristine one. While states 
near the Fermi level have mostly Fe $3d$ character, some of them also have 
appreciable Se-atom weight as denoted by the thickness of green lines in 
Figs.~\hyperref[fig2]{2(a)} and \hyperref[fig2]{2(b)}. This Se-atom weight 
reflects Se-mediated indirect hopping of Fe $3d$ electrons. These states 
with larger Se weights are lowered more in energy after K dosing because 
of the K-induced change of the electrostatic potential energy at the topmost Se 
atomic sites, denoted by $-\Delta$ in Fig.~\hyperref[fig1]{1(c)},
in addition to overall potential energy change by $-\delta\mu$. 
Especially, the hole 
band having the largest Se weight among the three hole bands at $\Gamma$ 
is so sensitive to K dosing that its maximum moves even below 
the Fermi level. As for electron bands near $M$, we observe a gradual decrease 
of bandwidths and increase of LDA effective masses ($m^{\text{LDA}}$).
In particular, $m^{\text{LDA}}$ of an electron band along the 
$\Gamma$ direction is obtained by a quadratic curve fit as 
marked with red dashed lines in Figs.~\hyperref[fig2]{2(a)} and 
\hyperref[fig2]{2(b)}. As shown in Fig.~\hyperref[fig2]{2(e)}, obtained 
$m^{\text{LDA}}$ in a pristine ML is close to the electron mass ($m_e$) 
at vacuum and increases more than 40\% with K dosing of 
$n_{\text{K}}=0.5$.

For the detailed analysis of K-dosing effects, we performed tight-binding
(TB) calculations. First, we constructed a TB Hamiltonian by applying 
maximally localized Wannier functions \cite{Mostofi:2014} to the LDA band 
structure of pristine FeSe ML. Then, we added $\delta\mu$ and $\Delta$ 
to our TB Hamiltonian and obtained a TB band structure, which agrees well 
with LDA result of K-dosed FeSe ML. We found that while $\delta\mu$ shifts
band energies rigidly, reflecting the electron doping, an increase of 
$\Delta$ reduces bandwidths of Fe $3d$ bands and increases the effective 
mass of the electron band at the $M$ point (Fig.~S4 in Ref. \cite{supplemental}).
To understand why $\Delta$ affects the bandwidth, we need to focus on the 
hopping mechanism in FeSe. Because of the small spatial extent of Fe $3d$ 
orbitals, direct intersite hoppings are relatively weak. Instead,
the dominant hopping channel for Fe $3d$ orbitals is indirect hopping 
mediated by Se $4p$ orbitals \cite{Yin:2011}. Since the indirect hopping 
is the second-order process, its strength is inversely proportional 
to the energy difference between Fe $3d$ and Se $4p$ orbitals.
Thus, the increase of $\Delta$, which makes Fe $3d$ and Se $4p$ more 
separated in energy by lowering the Se $4p$ energy, results in 
effectively reduced hopping between Fe $3d$ orbitals.

\begin{figure} 
\includegraphics[width=8.6cm]{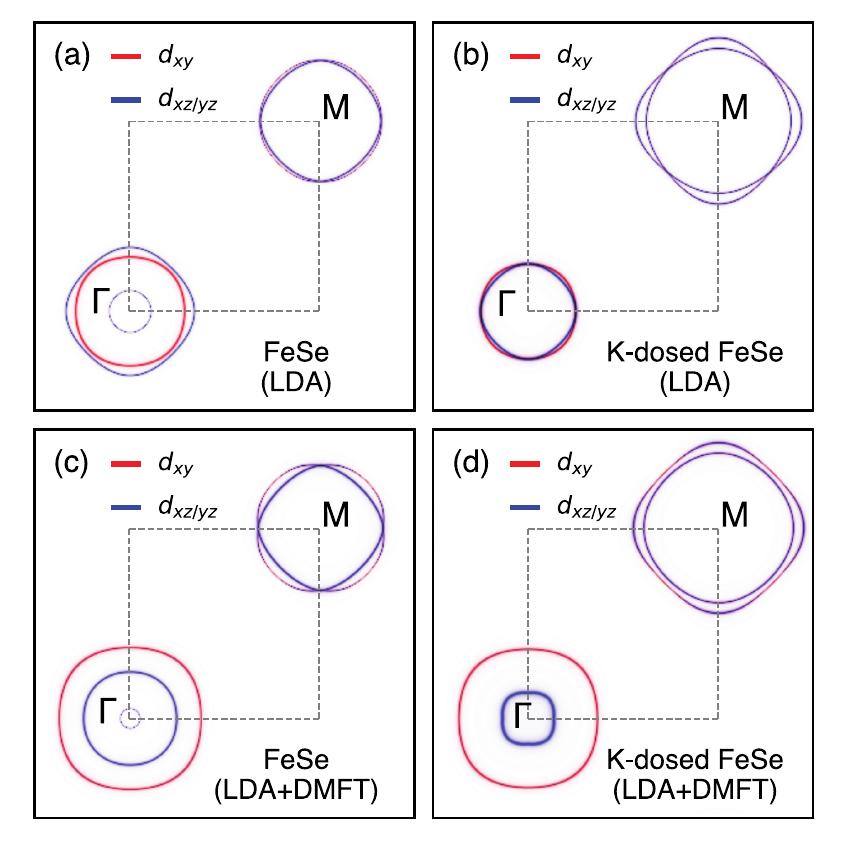}
\caption{\label{fig3} Fermi surface in FeSe ML.
(a),(b) LDA Fermi surfaces (a) without K dosing and 
(b) with $n_{\text{K}} = 0.3$. 
(c),(d) Zero-frequency LDA+DMFT spectral functions $A(\mathbf{k},\omega=0)$ 
for pristine and K-dosed cases. Blue and red colors represent 
$d_{xy}$ and $d_{xz/yz}$ orbital characters, respectively.
}
\end{figure}

We performed LDA+DMFT calculations for pristine and K-dosed FeSe MLs, and
obtained electronic structures as shown in Figs.~\hyperref[fig2]{2(c)} 
and \hyperref[fig2]{(d)}, respectively. We notice the electron correlation 
strongly renormalizes LDA band structures so that the bandwidths are shrunk
by more than a factor of 3 in our LDA+DMFT calculations. Because of the 
reduced hopping between Fe $d$ orbitals described above, the electron 
correlation becomes much stronger for K-dosed FeSe ML. For the direct 
comparison with LDA results, we fit the maxima of the LDA+DMFT spectral 
functions [red lines in Figs.~\hyperref[fig2]{2(c)} and 
\hyperref[fig2]{2(d)}] to extract the LDA+DMFT effective mass 
($m^{\text{LDA+DMFT}}$). Figure~\hyperref[fig2]{2(e)} shows that 
$m^{\text{LDA+DMFT}}$ increases over 60\% as $n_{\text{K}}$ increases 
to 0.5, and these values are comparable to the experimental 
observations \cite{Wen:2016,Seo:2016}.

The strength of electron correlation can be quantified with the DMFT mass 
enhancement factor ($Z^{-1}$), which is the inverse of the quasiparticle 
weight,
$Z^{-1} = 1-\left.\frac{\partial \text{Re}\Sigma}{\partial\omega}
\right|_{\omega\rightarrow 0}$, where $\Sigma$ is the DMFT self-energy.
In Fig.~\hyperref[fig2]{2(f)}, we show the orbital-resolved DMFT mass 
enhancement factors for K-dosed FeSe ML increase with $n_{\text{K}}$,
indicating that the K dosing makes FeSe more correlated. 
We notice $t_{2g}$ orbitals are more affected by 
K dosing and, especially, $d_{xy}$ electrons come to have an extremely strong
correlation, with $Z^{-1} > 5$ for $n_{\text{K}}=0.5$.

Figure~\hyperref[fig3]{3} shows how the K dosing and the electron correlation
modify the Fermi surface of FeSe ML.
Figures~\hyperref[fig3]{3(a)} and \hyperref[fig3]{3(b)} show the LDA Fermi 
surfaces before and after K dosing, respectively.
Two of the three hole pockets around $\Gamma$ are of $d_{xz/yz}$ 
characters and the other one is of $d_{xy}$. We note that the innermost 
hole pocket has appreciable weights from Se $4p$ orbitals  
so that it largely shifts below the Fermi level after K dosing.
Figures~\hyperref[fig3]{3(c)} and \hyperref[fig3]{3(d)} show the LDA+DMFT Fermi 
surfaces before and after K dosing, respectively.
Although electron correlation does not alter the number of Fermi-surface 
pockets, it distinguishes $d_{xy}$ and $d_{xz/yz}$ orbitals,
that is, the $d_{xy}$ hole pocket expands, while $d_{xz/yz}$ hole pockets 
shrink due to the electron correlation.
Two electron pockets around $M$ have mixed orbital characters of $d_{xy}$ 
and $d_{xz/yz}$. Upon K dosing, 
the size of electron pockets is enlarged as a result of the combined 
effects of both the electron doping and the increased effective masses.

\begin{figure} 
\includegraphics[width=8.6cm]{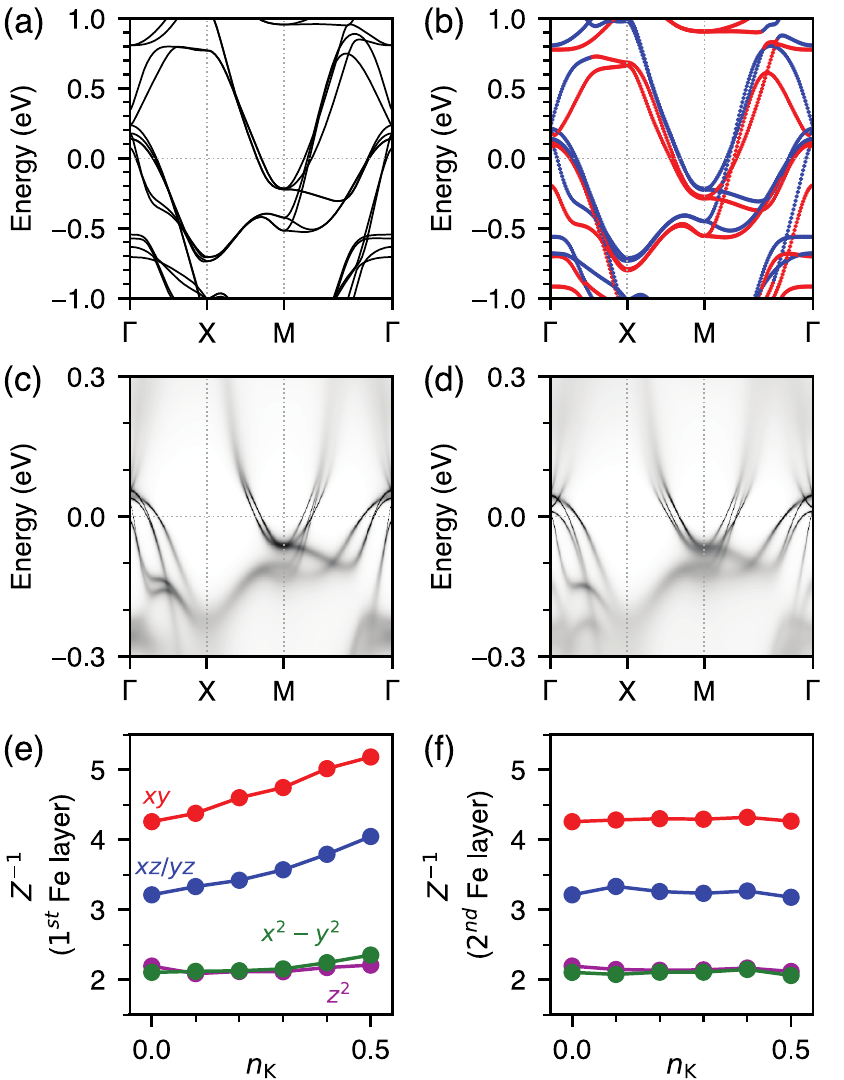}
\caption{\label{fig4} Electronic structure of FeSe BL.
(a),(b) LDA bands (a) without K dosing and (b) with $n_{\text{K}}=0.3$. 
Red (blue) circles are states with more weights in the first (second) layer.
(c),(d) LDA+DMFT spectral functions (c) without K dosing and
(d) with $n_{\text{K}}=0.3$.
(e),(f) DMFT mass enhancement factors of Fe $3d$ orbitals 
in the first and second FeSe layer.}
\end{figure}

Now, we consider the FeSe bilayer. Figures~\hyperref[fig4]{4(a)} and 
\hyperref[fig4]{4(b)} show our LDA band structure calculations of 
pristine FeSe BL and K-dosed FeSe BL with $n_{\text{K}}=0.3$, 
respectively. The band structure of pristine FeSe BL 
[Fig.~\hyperref[fig4]{4(a)}] is qualitatively the same as two 
sets of monolayer bands split by small interlayer coupling. Before 
K dosing, all the states near the Fermi level have equal weights 
at both the first and second layer due to the crystal symmetry. 
However, they are distinguished after K dosing, as shown in  
Fig.~\hyperref[fig4]{4(b)}, where states with more weights in the 
first (second) FeSe layer are colored red (blue). We notice that 
the K dosing affects only the states from the first FeSe layer, 
consistent with previous DFT study \cite{Zheng:2016} and 
experiment \cite{Wen:2016}. Similarly to FeSe ML, the K dosing lowers 
the energy of first-layer states and reduces their bandwidths, compared 
with the pristine BL case. With LDA+DMFT [Figs.~\hyperref[fig4]{4(c)} 
and \hyperref[fig4]{4(d)}], 
band dispersions are strongly renormalized by electron correlation,
and states from the first and the second layer are split in energy after 
K dosing. As clearly shown in Figs.~\hyperref[fig4]{4(e)} and 
\hyperref[fig4]{4(f)}, the DMFT mass enhancements of Fe $3d$ orbitals 
at the first layer increase with $n_{\text{K}}$, while those at the 
second layer remain almost constant. 

\begin{figure} 
\includegraphics[width=8.6cm]{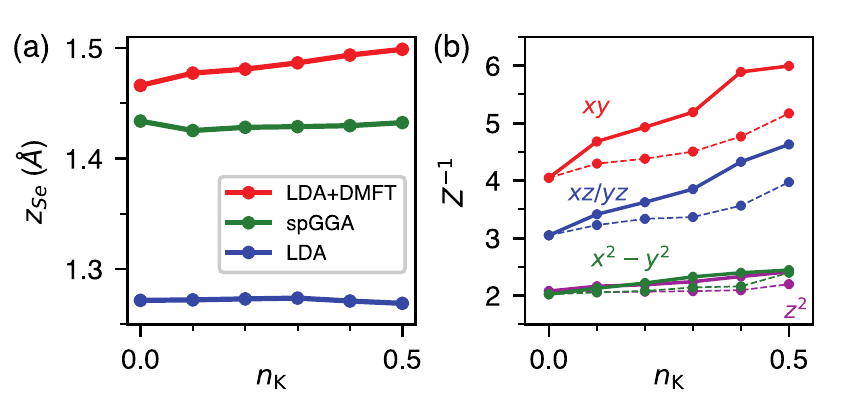}
\caption{\label{fig5} 
(a) Se heights ($z_{\text{Se}}$) in FeSe ML optimized at each
$n_{\text{K}}$ using different methods.
(b) DMFT mass enhancement factors ($Z^{-1}$) in FeSe ML. 
Solid lines are $Z^{-1}$ calculated with $z_{\text{Se}}$ optimized 
at each $n_{\text{K}}$ using LDA+DMFT. 
Dashed lines are $Z^{-1}$ with $z_{\text{Se}}$ fixed to
the pristine value, as shown in Fig.~\hyperref[fig2]{2(f)}.}
\end{figure}

So far we have focused on the direct electrical effects of K dosing on 
the electron correlation in FeSe layers by using the same atomic 
positions independently of $n_{\text{K}}$. Since the strength of 
electron correlation in FeSe is very sensitive to $z_{\text{Se}}$ 
\cite{Mandal:2017,Yin:2011,Haule:2016}, we investigate whether the K 
dosing can change $z_{\text{Se}}$ and thereby affect the electron 
correlation additionally. As shown in Fig.~\hyperref[fig5]{5(a)}, 
our LDA+DMFT calculations predict that the optimized 
$z_{\text{Se}}$ gradually increases with K dosing. While $z_{\text{Se}}$ 
is 1.47 {\AA} for the pristine case, it increases to 1.50 {\AA} with 
K dosing of $n_{\text{K}}=0.5$. Since higher $z_{\text{Se}}$ makes 
Fe $3d$ electrons more localized, the increase of $z_{\text{Se}}$ 
with K dosing enhances the electron correlation further as shown in 
Fig.~\hyperref[fig5]{5(b)}. We also find that the increase of 
$z_{\text{Se}}$ is mostly due to the enhanced electron correlation, 
which is captured by DMFT, rather than the direct electrostatic interaction 
between FeSe and K, because DFT calculations using LDA or the 
spin-polarized generalized gradient approximation predict 
nearly constant $z_{\text{Se}}$ insensitively to $n_{\text{K}}$.

In conclusion, our results show that K-dosed FeSe layers have stronger 
electron correlation than pristine ones because the change in
the electrostatic potential at the topmost Se atoms reduces Fe 3$d$ 
bandwidths by weakening the Se-mediated hopping. This enhancement of 
electron correlation, which occurs in the topmost FeSe layer only,
is indicated by the increased effective masses and reduced quasiparticle 
weights at the Fermi energy. Furthermore, the K dosing can increase the 
Se height, which enhances the electron correlation further. These results
shed a new light on comprehensive understanding of high-$T_c$ FeSe and 
can be generalized to other low-dimensional systems with surface, interface,
or gate doping.

\begin{acknowledgments}
This work was supported by NRF of Korea (Grant No.~2011-0018306). 
Y. W. C. acknowledges support from NRF of Korea 
(Global Ph.D. Fellowship Program NRF-2017H1A2A1042152).
Computational resources have been provided by KISTI
Supercomputing Center (Project No.~KSC-2017-C3-0079).
\end{acknowledgments}

\pagebreak
%%%%% SUPPLEMENTAL MATERIAL
\onecolumngrid
\setcounter{equation}{0}
\setcounter{figure}{0}
\setcounter{table}{0}
\renewcommand{\theequation}{S\arabic{equation}}
\renewcommand{\thefigure}{S\arabic{figure}}
\renewcommand{\bibnumfmt}[1]{[S#1]}
\renewcommand{\citenumfont}[1]{S#1}

\begin{center}

\textbf{Supplemental Material: \\
Role of Electric Fields on Enhanced Electron Correlation 
in Surface-Doped FeSe }\\[.2cm]
Young Woo Choi and Hyoung Joon Choi$^*$\\[.1cm]
{\itshape Department of Physics, Yonsei University, Seoul 03722, Korea\\}
(Dated: \today)
\end{center}

\begin{center}
\setlength{\fboxrule}{0pt}

\fbox{\begin{minipage}{0.8\textwidth}
    \hspace{5pt} This supplemental material provides (i) comparison of our virtual
crystal approximation of K-dosed FeSe layers and corresponding
ordinary supercell calculations, (ii) analysis of electron transfer,
electrostatic potential-energy change, and effective electric field
induced by K dosing, and (iii) tight-binding analysis of K-dosing
effects.
\end{minipage}}
\end{center}

\vspace{.1cm}
\twocolumngrid

\subsection{S1. Virtual crystal approximation for K dosing}

In our present work, we define the K concentration $n_{\text{K}}$
as the number of K atoms per surface unit cell which contains two
Fe atoms in each FeSe layer. We call this surface unit cell 
as 2-Fe unit cell hereafter. To simulate K-dosed FeSe layers with 
$n_{\text{K}} < 1$, we use a virtual crystal approximation (VCA) 
where one K atom is introduced to the 2-Fe unit cell and the atomic number 
of the K atom is reduced to $18+n_{\text{K}}$. 

To check the validity of our VCA method, we compare the VCA electronic 
structures of 
$n_{\text{K}}=0.25$ and $0.5$ cases with ordinary supercell calculations.
For $n_{\text{K}}=0.25$, we introduced one K atom to a $2\times 2$ supercell
of FeSe monolayer (ML) and performed ordinary LDA supercell calculation.
For clearer comparison, we unfolded the supercell band structure
[1--3] and obtained the unfolded LDA band structure as shown 
in Fig.~S1(a). For $n_{\text{K}}=0.5$, we introduced one K atom to 
a $\sqrt{2}\times\sqrt{2}$ supercell and obtained the unfolded LDA band 
structure as shown in Fig.~S1(b). Figures~S1(c) and (d) show our VCA band 
structures for $n_{\text{K}}=0.25$ and $0.5$, respectively. Except for some 
features arising from the broken translational symmetry in the supercell 
calculations, our VCA results agree excellently with the supercell
results. This supports the validity of our VCA method for K-dosed FeSe layers.

For $n_{\text{K}}>0.5$, a potassium band becomes so dispersive that it 
reaches down below the Fermi level, creating a new electron pocket near 
$\Gamma$ point. We do not regard this potassium band as fully relevant 
in a real sample since K atoms are likely to be disordered on the surface 
and may not form a coherent band. Thus, we limit our present work in
the range of $n_{\text{K}}\leq 0.5$

\begin{figure}
\includegraphics[width=9.0cm]{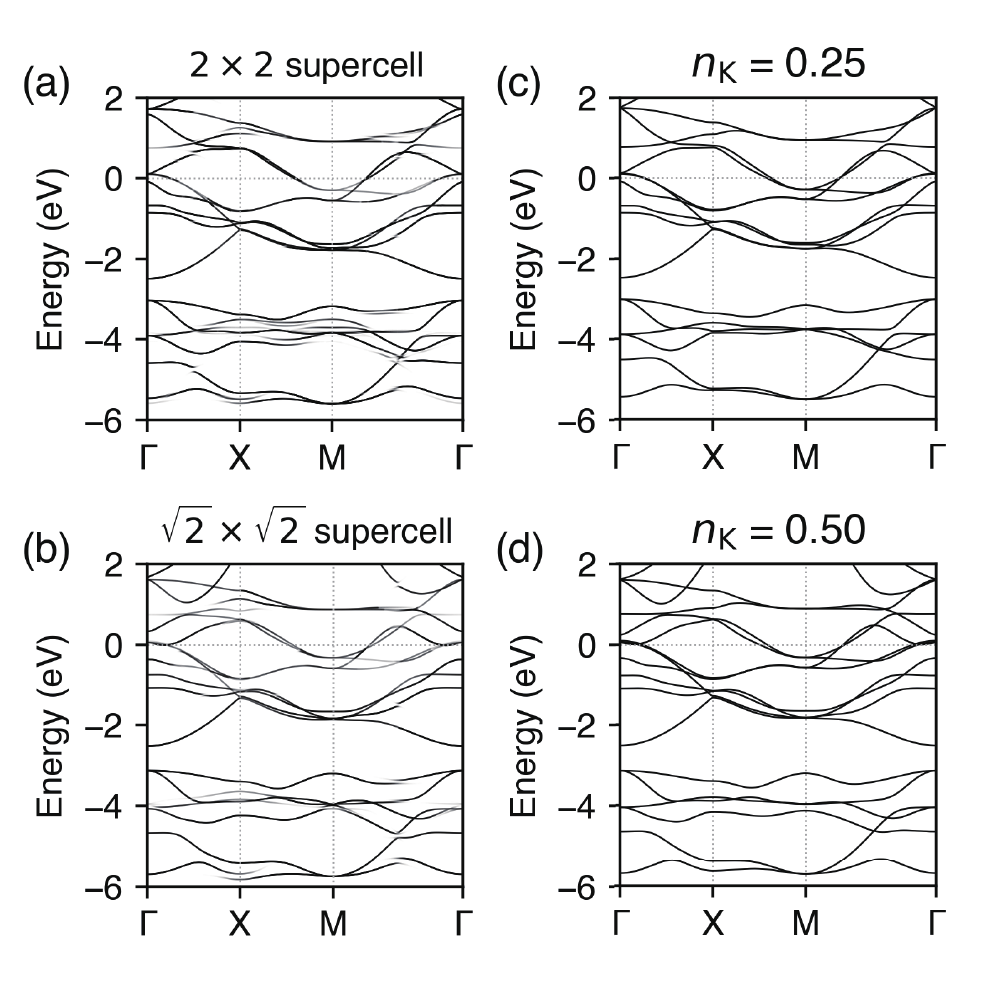}
\caption{Comparison of supercell and VCA results
for K-dosed FeSe monolayer.
(a) Unfolded LDA band structure from the supercell calculation with one K 
atom per $2\times 2$ supercell corresponding to $n_{\text{K}} = 0.25$.
(b) Unfolded LDA band structure from the supercell calculation with one K 
atom per $\sqrt{2}\times\sqrt{2}$ supercell corresponding 
to $n_{\text{K}} = 0.5$.
(c) VCA band structure with $n_{\text{K}}=0.25$.
(d) VCA band structure with $n_{\text{K}}=0.5$.
In (a)-(d), $\Gamma$, X, and M are high-symmetry points of the unit-cell 
Brillouin zone.
    }
\end{figure}

\subsection{S2. Electrostatic change after K dosing}

From self-consistent LDA calculations of FeSe bilayer (BL) with and without K 
dosing, we analyzed (i) the electron transfer ($\Delta\rho$) from K to FeSe, 
(ii) change ($\Delta V_{KS}$) in the Kohn-Sham effective potential before and 
after K dosing, and (iii) the effective electric field 
($-\partial(\Delta V_{KS})/\partial z$) generated by K dosing. For these 
analyses, we used the pseudopotential method as implemented in 
SIESTA [4]. For the K-dosed case, we used $n_{\text{K}} = 1$.

In more detail, we define the electron transfer $\Delta\rho$ as
$\Delta \rho = \rho^{\text{K-dosed FeSe BL}} - \rho^{\text{FeSe BL}} - 
\rho^{\text{K}}$, where $\rho^{\text{K-dosed FeSe BL}}$, 
$\rho^{\text{FeSe BL}}$, and $\rho^{\text{K}}$ are electron distributions 
in K-dosed FeSe BL, pristine FeSe BL, and one isolated K atom in the same 
surface unit cell, respectively. Figure~S2 shows $\Delta\rho$ averaged 
in the $xy$ plane and plotted along the $z$-direction. This plot shows that 
the electron is transferred from the K atom to the topmost FeSe layer, 
with negligible electron transfer to the second layer.

We define the change ($\Delta V_{KS}$) in the Kohn-Sham effective 
potential before and after K dosing as $\Delta V_{KS}=
V_{KS}^{\text{K-dosed FeSe BL}}-V_{KS}^{\text{FeSe BL}}$, where 
$V_{KS}^{\text{K-dosed FeSe BL}}$ and $V_{KS}^{\text{FeSe BL}}$ are 
self-consistent Kohn-Sham effective potentials in K-dosed and pristine
FeSe BLs, respectively. Figure~S2 shows $\Delta V_{KS}$ averaged in 
the $xy$ plane and plotted along the $z$-direction. We notice that 
the K dosing generates significant potential-energy difference 
between Fe and Se atom sites in the topmost FeSe layer, while
it slightly lowers the potential energy of Fe atom sites in the topmost 
FeSe layer.

Finally, we consider $-\partial(\Delta V_{KS})/\partial z$ as
indicating the effective electric field generated by K dosing.
Figure~S2 shows $-\partial(\Delta V_{KS})/\partial z$ averaged in
the $xy$ plane and plotted along the $z$-direction. We notice that 
a strong electric field is generated near the topmost surface Se atom site.

\begin{figure}
\includegraphics[width=6.5cm]{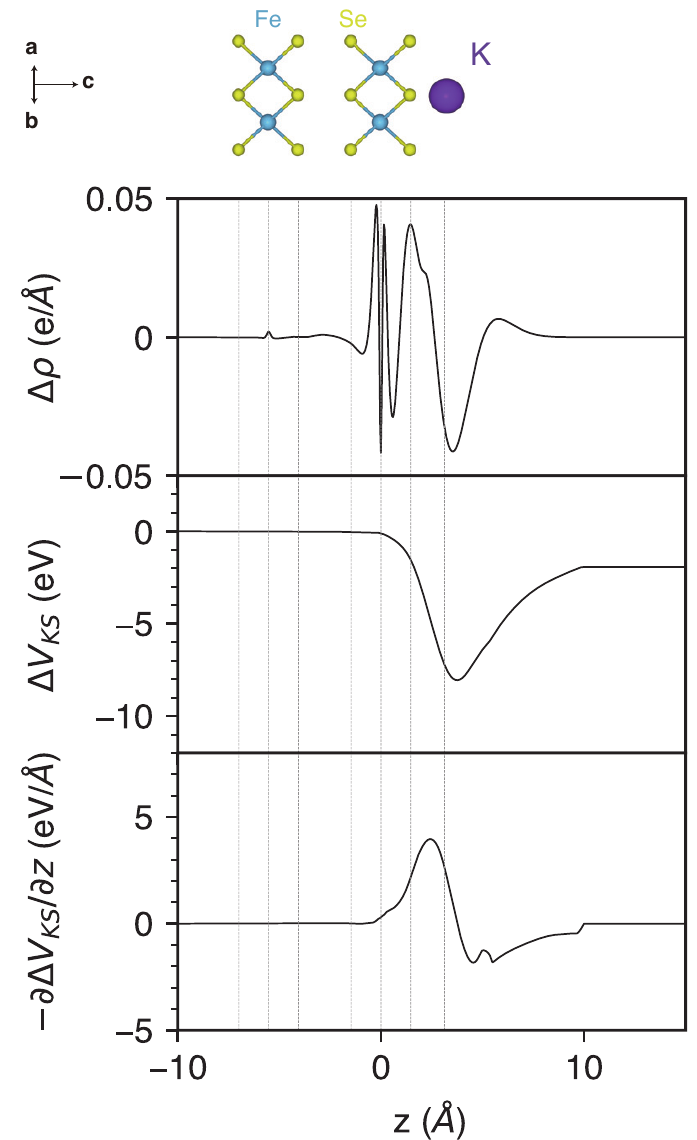}
\caption{ 
Electron transfer (top), potential-energy change (middle), and 
effective electric field (bottom) induced by K dosing, obtained 
by comparing LDA results of pristine FeSe BL and K-dosed FeSe BL with
$n_{\text{K}}=1$. Vertical lines indicate positions of atomic sites 
matching with the ball-and-stick model of K-dosed FeSe BL shown on top
of the plot. Definitions of $\Delta\rho$, $\Delta V_{KS}$, and
$-\partial(\Delta V_{KS})/\partial z$ are given in the text.
    }
\end{figure}

In order to show more clearly that only the first FeSe layer is affected
by dosed K atoms, we also considered FeSe four-layer instead of FeSe bilayer. 
As shown in Fig.~S3, only the topmost FeSe layer closest to K atoms is 
affected as in the case of FeSe bilayer 
while all the other three FeSe layers are almost unaffected.

\begin{figure}
\includegraphics[width=6.3cm]{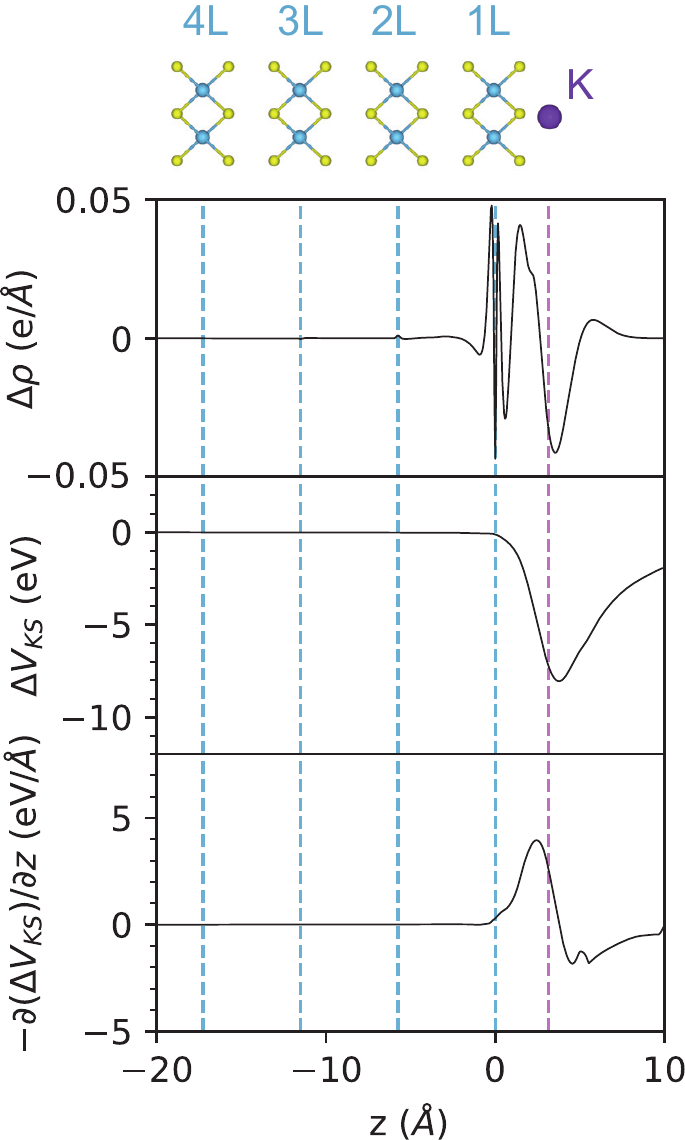}
\caption{ 
Electron transfer (top), potential-energy change (middle), and 
effective electric field (bottom) induced by K dosing, obtained 
by comparing LDA results of pristine FeSe four-layer and K-dosed FeSe four-layer with
$n_{\text{K}}=1$. 
$\Delta\rho$, $\Delta V_{KS}$, and
$-\partial(\Delta V_{KS})/\partial z$ are defined similarly to the bilayer case.
Vertical dashed lines indicate positions of Fe and K atomic sites 
matching with the ball-and-stick model of K-dosed FeSe four-layer shown on top
of the plot. 
    }
\end{figure}

\subsection{S3. Tight-binding analysis of K-dosing effects}

\begin{figure}
\includegraphics[width=8.6cm]{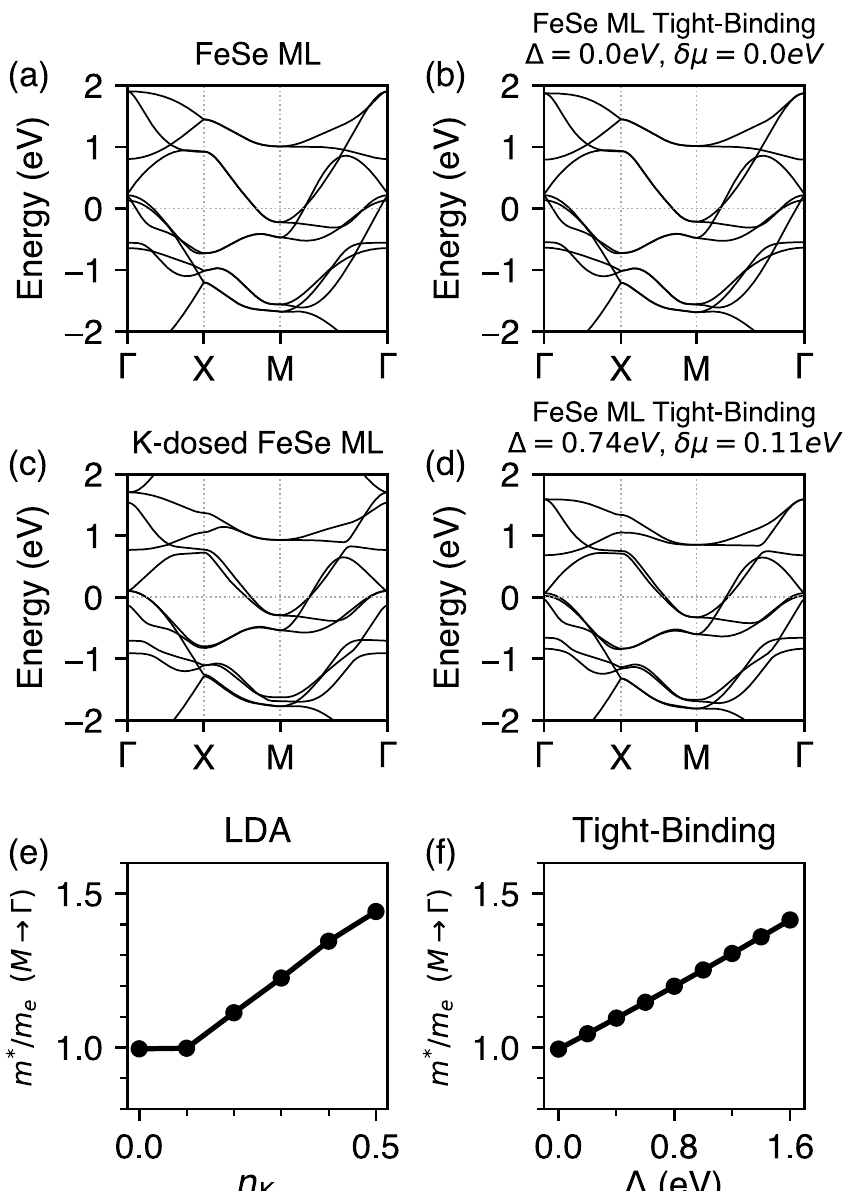}
\caption{Comparison of LDA and TB electronic structures.
(a) LDA bands for pristine FeSe ML. (b) TB bands for pristine FeSe ML.
(c) LDA bands for K-dosed FeSe ML with $n_{\text{K}} = 0.3$. (d) TB bands 
with $\delta\mu = 0.11$ eV and $\Delta = 0.74$ eV which correspond to 
$n_{\text{K}} = 0.3$. (e) LDA effective mass of an electron band along 
M$\rightarrow$$\Gamma$ direction as a function of the K concentration 
($n_{\text{K}}$). (f) TB effective mass of the corresponding electron band 
along M$\rightarrow$$\Gamma$ direction as a function of $\Delta$. 
For FeSe ML, $\delta\mu$ shifts the whole TB band structure rigidly.
}
\end{figure}

We construct a tight-binding (TB) model including all Fe $3d$ and Se $4p$ 
orbitals using the maximally localized Wannier functions [5, 6].
Figures~S4(a) and (b) show LDA and TB band structures of the pristine FeSe ML,
respectively. The TB bands agree well with the LDA results. This supports 
the validity of our TB model.

In K-dosed FeSe layers, effects of K dosing can be described 
by two changes in the TB Hamiltonian. One is the overall lowering 
(by $\delta\mu$) of the potential energy in the first FeSe layer with respect 
to the second FeSe layer, and the other is the additional lowering 
(by $\Delta$) of the potential energy at the topmost Se atom site with 
respect to the Fe atom site in the first FeSe layer. 
These two parameters, $\delta\mu$ and $\Delta$, capture 
essential features of the electronic structure of K-dosed FeSe systems. 
The lowering $\delta\mu$ brings surface electron doping to the system, 
and the lowering $\Delta$ accounts for the Fe bandwidth reduction 
by weakening Se-mediated hopping between Fe orbitals.

Figure~S4(c) shows LDA band structure for K-dosed FeSe ML with
$n_{\text{K}}=0.3$. This LDA band structure can be reproduced by varying 
$\delta\mu$ and $\Delta$ in our TB model [Fig.~S4(d)].
As clearly shown in Fig.~S4(f), increase of $\Delta$ increases the effective
mass of the Fe $3d$ band. This is because the main hopping channel of 
Fe $3d$ orbitals is the indirect hopping mediated by Se $4p$ orbitals
and $\Delta$ increases the energy separation between Fe $3d$ and Se $4p$
orbitals, weakening the indirect hopping.

\vspace{.7cm}

\noindent
$^*$ h.j.choi@yonsei.ac.kr

\vspace{.2cm}

\hangindent=.5cm\hangafter=1\noindent
[1] S. Kim, J. Ihm, H. J. Choi, and Y.-W. Son,
Origin of anomalous electronic structures of epitaxial graphene on 
silicon carbide, Phys. Rev. Lett. {\bf 100}, 176802 (2008).

\hangindent=.5cm\hangafter=1\noindent
[2] H. Lee, S. Kim, J. Ihm, Y.-W. Son, and H. J. Choi,
Field-induced recovery of massless Dirac fermions in epitaxial graphene on SiC,
Carbon {\bf 49}, 2300 (2011).

\hangindent=.5cm\hangafter=1\noindent
[3] O. Rubel, A. Bokhanchuk, S. J. Ahmed, and E. Assmann, 
Unfolding the band structure of disordered solids: From bound states to high-mobility Kane fermions, 
Phys. Rev. B {\bf 90}, 115202 (2014).

\hangindent=.5cm\hangafter=1\noindent
[4] J. M. Soler, E. Artacho, J. D. Gale, A. Garca, 
J. Junquera, P. Ordejn, and D. Snchez-Portal, 
The SIESTA method for ab initio order- N materials simulation, 
J. Phys.: Condens. Matter {\bf 14}, 2745 (2002).

\hangindent=.5cm\hangafter=1\noindent
[5] A. A. Mostofi, J. R. Yates, G. Pizzi, Y.-S. Lee, 
I. Souza, D. Vanderbilt, and N. Marzari, 
An updated version of wannier90: A tool for obtaining 
maximally-localised Wannier functions, 
Comp. Phys. Commun. {\bf 185}, 2309 (2014).

\hangindent=.5cm\hangafter=1\noindent
[6] J. Kune, R. Arita, P. Wissgott, A. Toschi, H. Ikeda, and K. Held, 
Wien2wannier: From linearized augmented plane waves to maximally 
localized Wannier functions, 
Comp. Phys. Commun. {\bf 181}, 1888 (2010).


\begin{thebibliography}{99}
\bibitem{note}
$K$ dosing also causes tiny lateral displacements (less than 0.01~{\AA}) 
of Fe and Se atoms in the first FeSe layer only, which
have negligible effect on the electronic structure.
\end{thebibliography}
\end{document}